\documentclass[aps,english,prb,reprint,superscriptaddress, 
citeautoscript,showpacs]{revtex4-1}
\usepackage{graphicx}
\usepackage{amssymb}
\usepackage{babel}
\usepackage{color}
\usepackage{multirow}
\usepackage{setspace}
\usepackage{longtable}
\begin{document}

\title{MoSi$_{2}$N$_{4}$ single-layer: a novel two-dimensional material with outstanding mechanical, thermal, electronic, optical, and photocatalytic properties}

\author{A. Bafekry}\email{bafekry.asad@gmail.com}
\affiliation{Department of Physics, University of Guilan, 41335-1914 Rasht, Iran}
\affiliation{Department of Physics, University of Antwerp, Groenenborgerlaan 171, B-2020 Antwerp, Belgium}
\author{M. Faraji}
\affiliation{Micro and Nanotechnology Graduate Program, TOBB University of Economics and Technology, Sogutozu Caddesi No 43 Sogutozu, 06560, Ankara, Turkey}
\author{Do M. Hoat}
\affiliation{Computational Laboratory for Advanced Materials and Structures, Advanced Institute of Materials Science, Ton Duc Thang University, Ho Chi Minh City, Vietnam}
\author{M. M. Fadlallah}
\affiliation{Department of Physics, Faculty of Science, Benha University, 13518 Benha, Egypt}
\author{M. Shahrokhi}
\affiliation{University Lyon, ENS de Lyon, CNRS, Université Claude Bernard Lyon 1, Laboratoire de Chimie UMR 5182, F-69342 Lyon, France}
\author{F. Shojaei}
\affiliation{Department of chemistry, Faculty of sciences, Persian gulf university, Bushehr 75169, Iran}
\author{D. Gogova} 
\affiliation{Department of Physics, University of Oslo, P.O. Box 1048, Blindern, Oslo, Norway}
\author{M. Ghergherehchi}\email{mitragh@skku.edu}
\affiliation{College of Electronic and Electrical Engineering, Sungkyun kwan University, Suwon, Korea}

\begin{abstract} 
Very recently, the two-dimensional (2D) form of MoSi$_{2}$N$_{4}$ has been successfully fabricated [Hong et al., Sci. 369, 670 (2020)]. Motivated by theses recent experimental results, herein we investigate the structural, mechanical, thermal, electronic, optical and photocatalytic properties using hybrid density functional theory (HSE06-DFT). Phonon band dispersion calculations reveal the dynamical stability of MoSi$_{2}$N$_{4}$ monolayer structure. Furthermore, the mechanical study confirms the stability of MoSi$_{2}$N$_{4}$ monolayer. As compared to the corresponding value of graphene, we find the Young's modulus decreases by $\sim$ 30\% while the Poisson's ratio increases by $\sim$ 30\%. 
In addition, its work function is very similar to that of phosphorene and MoS$_{2}$ monolayers. The electronic structure investigation shows the MoSi$_{2}$N$_{4}$ monolayer is an indirect bandgap semiconductor. We have determined the bandgap using the HSE06 (GGA) is 2.35 (1.79) eV, which is an overestimated (underestimated) value of the experimental bandgap (1.99 eV).
The thermoelectric study shows a good thermoelectric performance of the MoSi$_{2}$N$_{4}$ monolayer with a figure of merit slightly larger than unity at high temperatures. The optical analysis using the RPA method constructed over HSE06 shows that the first absorption peak of the MoSi$_{2}$N$_{4}$ monolayer for in-plane polarization is located in the visible range of spectrum, i.e. it is a promising candidate for advancing optoelectronic nanodevices. The photocatalytic study indicates the MoSi$_{2}$N$_{4}$ monloayer can be a promising photocatalyst for water splitting as well as and CO$_{2}$ reduction. In summary, the fascinating MoSi$_{2}$N$_{4}$ monloayer is a promising 2D material in many applications due to its unique physical properties.
\end{abstract}

\maketitle

\section{Introduction}

In recent years, 2D compounds are of interest because of their unique properties, including high surface to volume ratio, high ionicity, low phonon energy, and so on. The 2D graphene-like materials are the most significant structures due to their mechanical, thermal, electronic, optical, and thermoelectrical properties. They can either be nanostructured materials representing metallic \cite{Monika2020}, half-metallic \cite{Chilukuri}, semiconducting \cite{Mostafa2020}, insulating \cite{Yin2016}, or superconducting \cite{Yoshida}. 
Therefore, they are an intriguing topic for experimental and theoretical studies due to their promising candidates for many potential applications in electronic and optoelectronics \cite{Kang2020}, thin-film field-effect transistors \cite{Goki2008}, catalysis \cite{Huang2017}, energy storage \cite{Bonaccorso2015} such as Li-ion batteries \cite{Eunha2018} and supercapacitors \cite{Patrice2008}. The 2D materials can be synthesized by wet chemistry \cite{wet}, hydrothermal \cite{hydro}, 
magnetron sputtering \cite{magneto}, atomic layer deposition \cite{atomic}
physical vapor deposition (PVD) \cite{PVD}, and chemical vapor deposition (CVD) \cite{CVD1,CVD2}, and so on. But they are unstable in the presence of water and oxygen. Recently, a very high-quality 2D transition metal nitride and carbide synthesized using CVD with diverse hexagonal, cubic, and orthorhombic crystalline structures. Yi-Lun Hong et al. \cite{Hong2020} synthesized MoSi$_2$N$_4$ using CVD with NH$_3$ gas as the source of nitrogen, a Cu/Mo bilayer as the substrate, and elemental Si, in triangular shapes and uniform thickness. They also employed first-principle density functional theory (DFT) to calculate the electronic properties of the synthesized MoSi$_2$N$_4$, in which the results depicted a direct bandgap semiconductor with an energy bandgap of 1.744 eV. J. Guo et al. \cite{Guo2015} reported strong heavy metal ion absorption on 2D alkalization intercalated Mxene (alk-Mxene) using DFT. They have exhibited that the occupation of the F atom decreases the ion absorption efficiency of alk-Mxene but it increases dramatically by the presence of K, Na, and Li atoms and the presence of hydroxyl group position perpendicular to Ti atoms, depicts the strongest tendency for eliminating the heavy metal ion. N. Shirchinnamjil et al. \cite{Nyamdelger2020} explored 2D Mxene V$_2$C monolayer potential as an anode for Li-ion batteries using DFT calculations. They have reported the diffusion mobility and storage capacity dependency of Li ions which can be strongly absorbed on the defective and non-defective of the two sides of the 2D V$_2$C structure. J. D. Gouveia et al. \cite{Gouveia} reported absorption of six amino acids on the 2D Ti$_2$CO$_2$. They found that most amino acids tend to adsorb with their N atom, with a very weak bond, and with a Ti atom, with a strong bond, while other amino acids prefer to adsorb to the Ti$_2$CO$_2$ surface parallelly and the van der Waals forces are dominated to all structures. Finally, it has been depicted that the Ti-N bond has a very weak character.

In this paper, we present the details of the calculation methods in section II. We investigate the stability of MoSi$_2$N$_4$ monolayer by dynamical (section III) and mechanical methods (section IV). We tackle the electronic structure using the hybrid exchange correlation functional (HSE06) in section V. Then we discuss the thermoelectric properties of MoSi$_2$N$_4$ monolayer by calculating the Seebeck coefficient, electrical conductivity, thermal conductivity, power factor and figure of merit in section VI. We study the optical and photocatalytic properties of MoSi$_2$N$_4$ monolayer in section VII . Finally, in section VIII, we present a summary of the main results and the conclusions.

\section{Method}
The plane-wave basis projector augmented wave as implemented in the Vi-
enna ab-initio Simulation Package (VASP) \cite{vasp1,vasp2}. was employed in the framework of DFT. The generalized gradient approximation (GGA) in the Perdew-Burke-Ernzerhof form \cite{GGA-PBE1,GGA-PBE2} and hybrid Heyd-Scuseria-Ernzerhof functiona (HSE06) \cite{Heyd} were used for the exchange-correlation potential. The kinetic energy cut-off of of 600 eV, and a $\Gamma$-centered 16$\times$16$\times$1 {\it k}-mesh for the unit cell were employed in our calculations. The tolerance of the total energy were converged to less than 10$^{-5}$ eV with forces less than 10$^{-3}$ eV \AA{}$^{-1}$. The lattice constants and atomic positions were optimized without any constraint. The vacuum space was $\sim$20 \AA{} along the \textit{z}-direction to avoid any fictitious interactions. The Bader charge analysis \cite{henkelman2006fast} was utilized to extract the electronic charge transfers. 

The vibrational properties were computed by the finite-displacement method implemented in the PHONOPY code\cite{alfe2009phon}. The optical spectra were performed in the random phase approximation \cite{Gajdos} method constructed over HSE06 using 
a dense {\it k}-grid of $20\times20\times1$. The wave function in the interstitial region was expanded in terms of plane waves with a cut-off parameter of $R_{MT}$ $K_{max}$= 8.5, where $R_{MT}$ and $K_{max}$ denote the smallest atomic sphere radius and the largest k vector, respectively (see supporting information for more details). The thermoelectric properties of the MoSi$_{2}$N$_{4}$ single layer was calculated by the interpolation of the band structure supported by the BoltzTraP\cite{BoltzTraP}. It is important to mention that the accuracy of thermoelectric properties is extremely sensitive to the band structure, therefore, we used a very dense {\it k}-mesh of $26\times26\times1$. 

\section{Structure and Stability}

\begin{figure}[!b]
\includegraphics[scale=1]{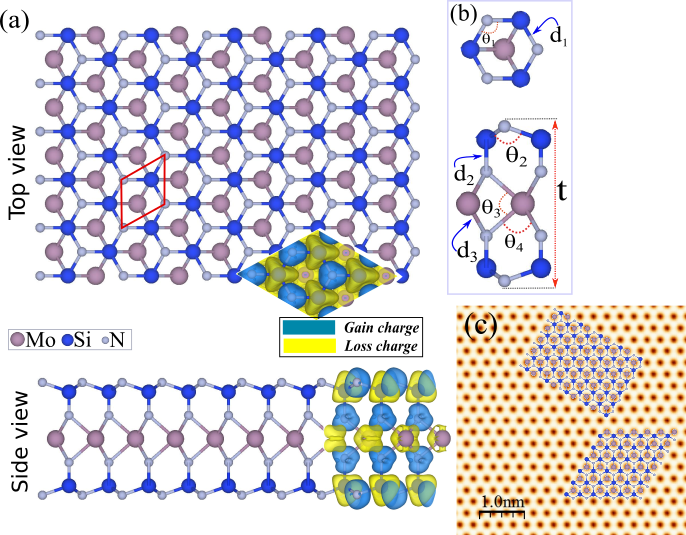}
\caption{(Color online) Top and side view of crystal structure of MoSi$_{2}$N$_{4}$ single layer. The primitive unit cell is indicated by a red hexagonal. Difference charge density is shown in the inset. (b) Schematic structure parameters and (c) Simulated STM image of MoSi$_{2}$N$_{4}$ monolayer. The inset structure represents repeating the unit cell.}
\label{1}
\end{figure}

\begin{figure}[!htb]
\includegraphics[scale=1]{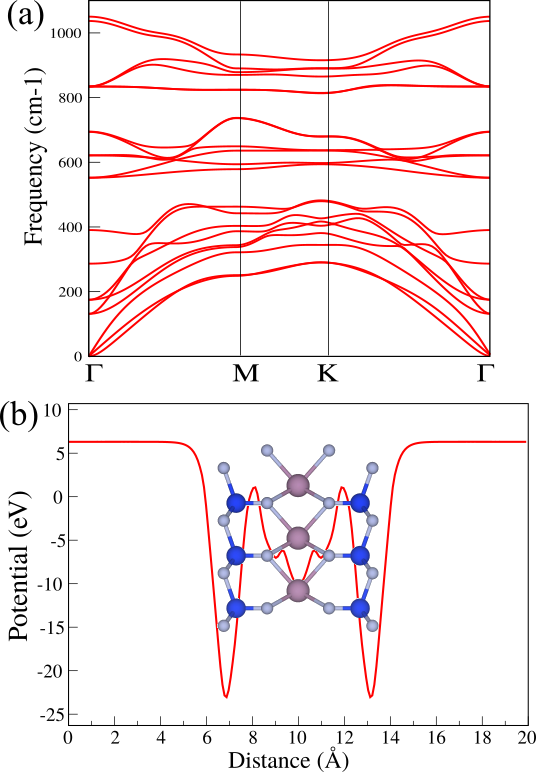}
\caption{(Color online) (a) Phonon band dispersion and (b) potential average of MoSi$_{2}$N$_{4}$ single layer.}
\label{2}
\end{figure}

\begin{table*}
\centering
\caption{\label{table1} The structural and electronic parameters including lattice constant $\textbf{a,b}$; 
Si-N ($d_{1,2}$) and Mo-N ($d_{3}$) bond lengths; 
Si-N-Si ($\theta_{1},_{2}$) and Si-Mo-Si ($\theta_{3},_{4}$) angles; 
the bucking of MoSi$_{2}$N$_{4}$ defined by the difference between the largest and smallest z coordinates of Zn and Sb atoms $(\Delta{z})$; 
the thickness layer $(t)$;
the cohesive energy per atom, $(E_{coh})$; 
the charge transfer $(\Delta{Q})$ between atoms;
the work function ($\Phi$);
the band gap $(E_{g})$ of PBE and HSE06 are shown outside and inside parentheses, respectively; VBM/CBM positions.
}
\begin{tabular}{lcccccccccccccc} 
\hline\hline
&\textit{a},\textit{b} & \textit{d$_{1,2}$}& \textit{d$_{3}$} &$t$ & \textit{$\theta_{1,2}$}&\textit{$\theta_{3,4}$}& $E_{coh}$ & $\Delta{Q_{1,2}}$ & $\Phi$ &$E_{g}$& VBM/CBM\\ 
& (\AA) & (\AA) & (\AA) & (\AA) & ($^{\circ}$) & ($^{\circ}$) & (eV/atom) & (e) & (eV) & (eV) & \\
\hline
MoSi$_{2}$N$_{4}$  & 2.91  & 1.75,1.74& 2.09 & 7.01 & 112,106 & 73,87 & 38.46 & 2.99,1.5 & 5.12 & 1.79 (2.35) & $\Gamma$/K\\
\hline\hline
\end{tabular}
\end{table*}
  
Top and side views of the crystal structure of MoSi$_{2}$N$_{4}$ monolayer are shown in Fig.\ 1(a). From the top view, the MoSi$_{2}$N$_{4}$, Mo, Si, and N atoms are packed in a honeycomb lattice, forming a 2D crystal with a space group of P6m1. The side view illustrates that this crystal is composed of covalently bonded atomic layers in the order of N-Si-N-Mo-N-Si-N along the c-axis with a thickness of 6 \AA{}. The monolayer can be viewed as a 2H-MoS$_{2}$-like MoN$_{2}$ layer sandwiched in-between two slightly buckled honeycomb SiN layers. The three layers are stacked on each other in a way that Mo atoms are located right below the center of Si$_{3}$N$_{3}$ hexagons of SiN layers and the two SiN monolayers are bonded to MoN$_{2}$ layer via Si-N bonds. 
The side view of MoSi$_{2}$N$_{4}$ showed that each single-layer block consisted of one layer of heavy atoms sandwiched by two layers of light atoms with a distance of $\sim$6 \AA{}.
The calculated lattice constants are equal to 2.91 \AA{}, while the bond lengths $d_{1}$ and $d_{2}$ (Si-N) are determined to be 1.75 and 1.74 \AA{}, respectively. 
Also the bond length of Mo-Si ($d_{3}$) is calculated 2.09 \AA{}. The two angles of Si-N-Si in lattice of MoSi$_{2}$N$_{4}$ are 112$^{\circ}$ ($\theta_{1}$) and 106$^{\circ}$ ($\theta_{2}$), while two angles of Si-Mo-Si are 87$^{\circ}$ ($\theta_{3}$) and 73$^{\circ}$ ($\theta_{4}$), which strongly deviating from 120$^{\circ}$ and indicating the out-plane anisotropy of the lattice. The thickness $(t)$ of MoSi$_{2}$N$_{4}$ monolayer is 7.01 \AA{}. These results are good agreement with a previous report \cite{Hong2020}.

The difference charge density of MoSi$_{2}$N$_{4}$ single layer is shown in the inset of Fig. \ref{1}, where the blue and yellow regions represent charge accumulation and depletion, respectively. The negatively charged N atoms are surrounded by positively charged Si and Mo atoms due to a charge transfer from Si and Mo atoms to N atom. 
The charge analysis shows that each N atom gains about 2.23 and 1.5 $e^-$ from the adjacent Si and Mo atom in MoSi$_{2}$N$_{4}$ single-layer, respectively. 

The cohesive energy $E_{coh}$ is then given by
\begin{equation}
E_{coh} = \frac{E_{Mo}+2E_{Si}+4E_{N}-E_{tot}}{n_{tot}},
\end{equation}
where $E_{Mo}$, $E_{Si}$ and $E_{N}$, $E_{tot}$ represent the energies of isolated Mo, Si, N atoms and total energy of the monolayer, n$_{tot}$ is the total number of unit cells, respectively. The cohesive energy is found to be -38.46 eV/atom. The negative energy indicates the stability of structure. 

The dynamical stability of MoSi$_{2}$N$_{4}$ single-layer is verified by calculating their phonon band dispersions through the whole Brillouin zone which is presented in Fig. \ref{2}(a). Phonon branches are free from any imaginary frequencies indicating the dynamical stability of the structure. The crystal structure of MoSi$_{2}$N$_{4}$ single-layer exhibit 3 acoustic and 16 optical phonon branches. Among 16 of the optical branches, 11 of them are found to be non-degenerate out-of-plane vibrational modes while the remaining 5 are three different doubly-degenerate phonon modes. The three acoustic branches, the frequency of the out-of-plane vibrational mode is quadratic in frequency as the frequencies tend to zero. The gap in the phonon dispersion separates the degenerate and no-degenerate of the optical branches even at the high-symmetry points at the zone-boundary. The almost dispersionless phonon at $\sim$580 cm$^{-1}$, 
610 cm$^{-1}$ and 820 cm$^{-1}$ is called homopolar mode, which is related to the layered
structures. The lattice vibrations corresponding to a change in the layer thickness. dispersionless phonon can be attributed to the light layers (SiN) vibrating
in counter phase in the normal direction, while the heavy layers (MoN) remain stationary \cite{phodis}.

Figure \ref{2}(b) shows the electrostatic potential for MoSi$_{2}$N$_{4}$ single layer. The potential is flat in the vacuum region and is symmetric around Mo atoms.
The calculated work function, $\Phi =E_{vacuum}-E_{F}$, is 5.12 eV which is similar to the corresponding values of phosphorene \cite{phos} and MoS$_{2}$ \cite{WMos} monolayers and larger than graphene (4.5 eV) \cite{Wgra}. 
 
\section{Mechanical properties}
The elastic constants of MoSi$_{2}$N$_{4}$ in the framework of harmonic approximation have three independent elastic constants of C$_{11}$, C$_{12}$, and C$_{66}$. The calculation of elastic parameters was discussed extensively in our previous work \cite{Asad1}. 
The MoSi$_{2}$N$_{4}$ is mechanically stable if it satisfies Born criteria \cite{Klepeis} of C$_{11}$ $>$ 0, C$_{11}$-C$_{12}$	$>$ 0, and C$_{66}$ $>$ 0, where C$_{11}$, C$_{12}$, and C$_{66}$ are linear elastic constants. According to our calculations C$_{11}$, C$_{12}$, and C$_{66}$ are found to be 533.99, 151.83, 1nd 100.80 N/m, respectively. Obviously these values satisfy the Bron criteria for the hexagonal lattice, confirming the mechanical stability of MoSi$_{2}$N$_{4}$. We also studied the mechanical properties of MoSi$_{2}$N$_{4}$. 
The mechanical properties by two independent parameters of in-plane stiffness ($C$) and Poisson$’$s ratio ($n$). 
The in-plane stiffness along X and Y directions are the same where C$_{x}$= C$_{y}$ = C$_{11}$ - C$_{12}$2/C$_{11}$. The stiffness value is 490.82 N/m. Assuming an approximate effective thickness for the monolayer, its Young's modulus is 0.6 TPa, which is almost one-third of the experimentally obtained value for graphene \cite{Jae-Ung}. The material also possesses same Poisson's ratio along $X$ and $Y$ direction and they are calculated using $\nu_{x}$ = $\nu_{x}$ = C$_{12}$/C$_{11}$. We find the Poisson$’$s ratio is 0.28 which is larger than that of graphene (0.16) \cite{Yue}.

\section{Electronic properties}

\begin{figure}[!b]
\includegraphics[scale=1]{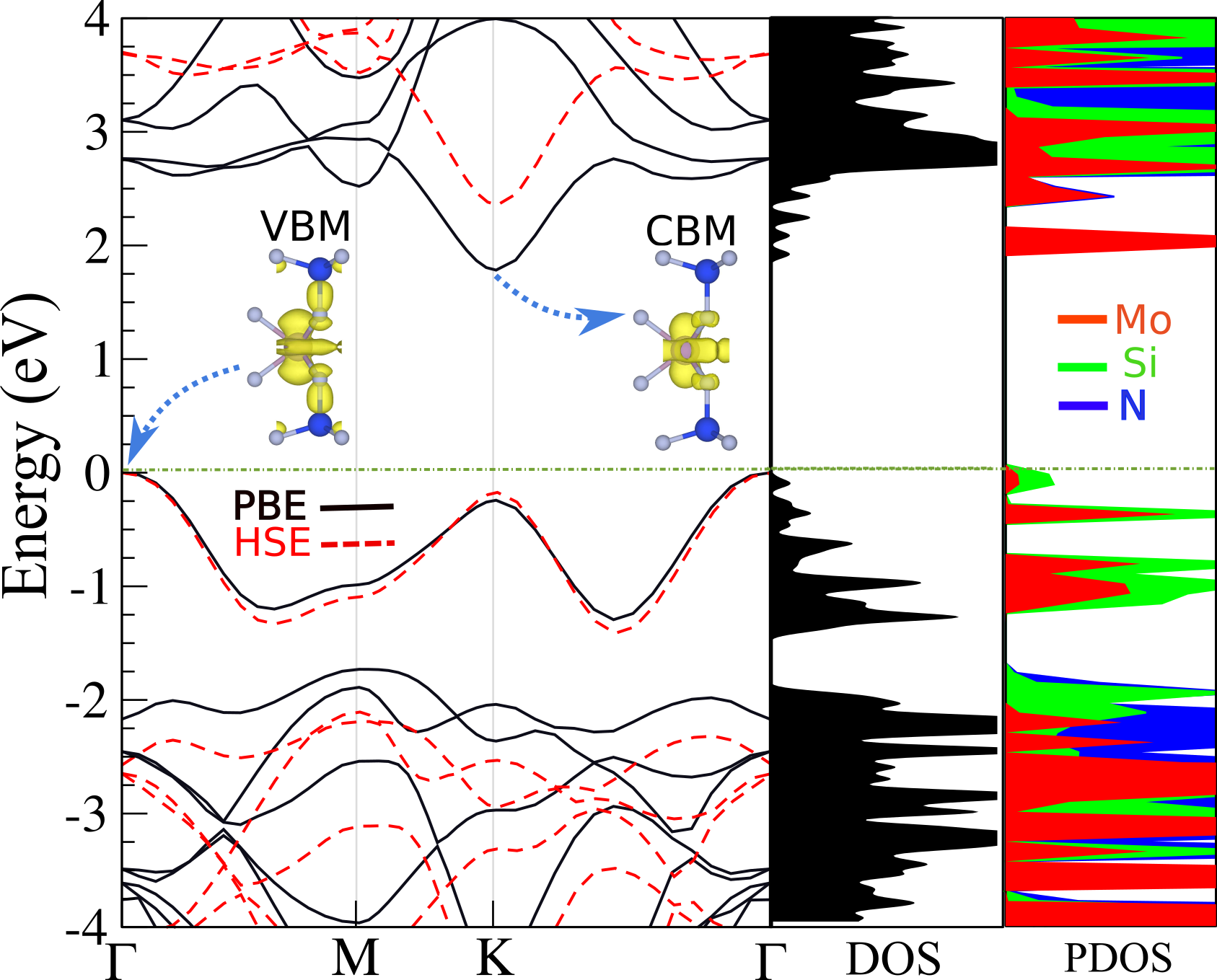}
\caption{Electronic band structure, DOS and PDOS of MoSi$_{2}$N$_{4}$ single-layer within PBE and HSE06 functionals. Charge densities of the valance band maximum (VBM) and conduction band minimum (CBM) orbitals are shown in the inset in band structure. The zero of energy is set to Fermi-level.}
\label{3}
\end{figure}

Next, we investigate the electronic properties of MoSi$_{2}$N$_{4}$ monolayer.
Fig.\ref{3} depicts the HSE06 and PBE electronic band structures, density of states (DOS), and partial DOS (PDOS) of MoSi$_{2}$N$_{4}$ monolayer. 
We find MoSi$_{2}$N$_{4}$ monolayer has an indirect gap semiconductor with an HSE06 (PBE) band gap of 2.35 (1.79) eV. The valence band maximum (VBM) is located at the $\Gamma$-point while that conduction band minimum (CBM) is located at the K-point. 
The lowest direct gap occurs at K-point and it is larger than the indirect gap by 0.17 eV. Both HSE06 and PBE band gap values and transition k-points are in a very good agreement with a previous report \cite{Hong2020}. 
In order to gain insight into the nature of VBM and CBM (see the inset of Fig.\ref{3}), the VBM is mainly constructed by $d_{z^{2}}$ and $d_{x^{2}-y^{2}}$ Mo states with a minor contribution of $p_{z}$ Si and N states, representing  $\sigma$(Mo-Mo) bonding hybridized with $\sigma$(N-Si) bonding (see Fig. S1 of the supplementary information (SI)). However, the CBM is solely contributed by $d_{z^{2}}$ Mo states representing a $\sigma$(Mo-Mo) bonding. The hole (electron) effective masses of the MoSi$_{2}$N$_{4}$ at $\Gamma$ $\longrightarrow$ M (K $\longrightarrow$ M) is 1.05 $m_{e}$$^*$ (0.45 $m_{e}$$^*$) while it is 1.16 $m_{e}$$^*$ (0.44 $m_{e}$$^*$). The light electron effective masses lead to the high carriers mobility in MoSi$_{2}$N$_{4}$ single-layer. 

\section{Optical properties}

\begin{figure}
	\includegraphics[scale=1]{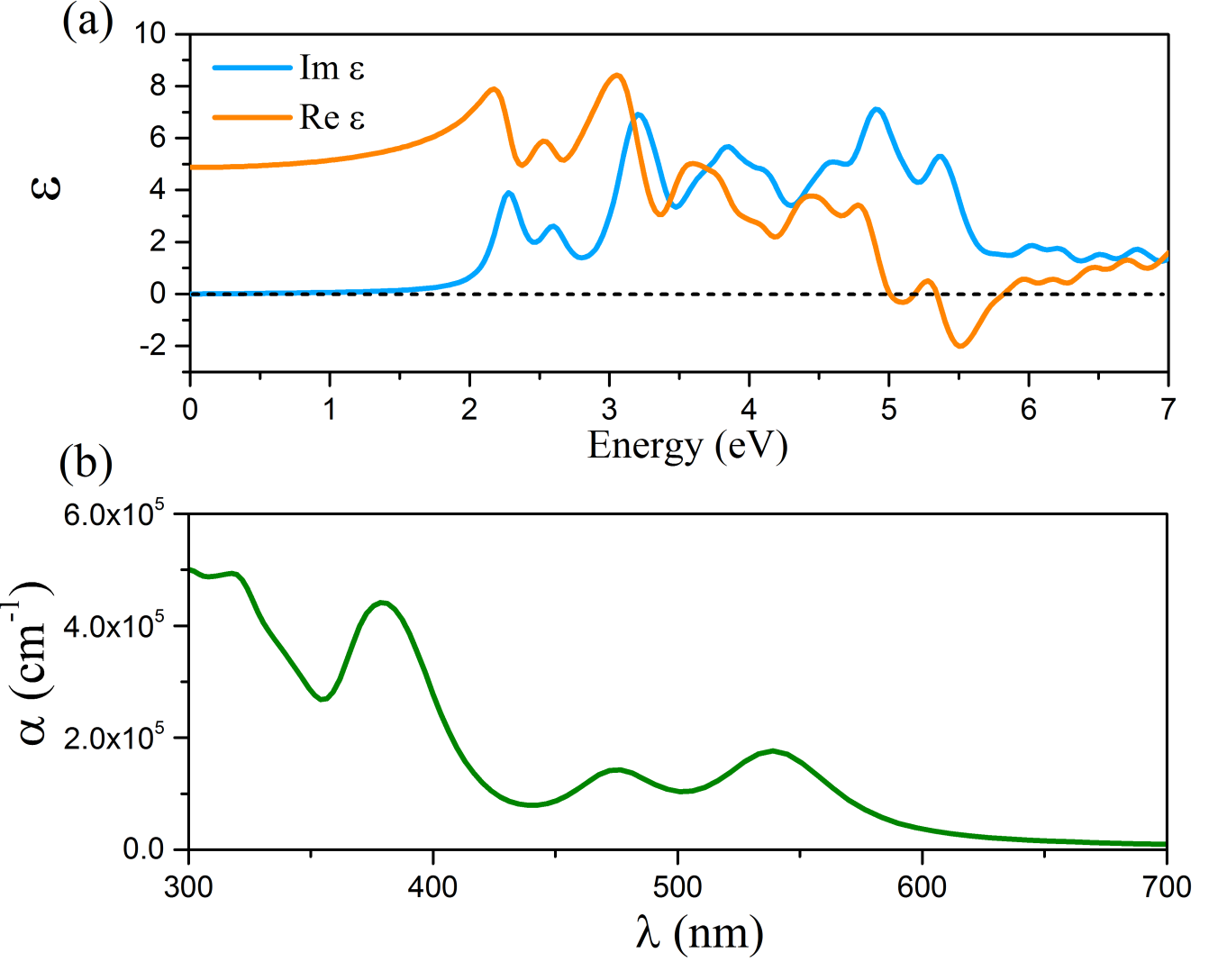}
	\caption{(a) Imaginary and real parts of the dielectric function as a function of photon energy and (b) optical absorption spectra as a function of wave length for MoSi$_{2}$N$_{4}$ monolayer along in-plane polarization, predicted using the RPA + HSE06 approaches.}
	\label{4}
\end{figure}

We now move to discuss the optical responses of this novel 2D system using the RPA method constructed over HSE06. Because of the symmetric geometry along the x- and y- axes the optical spectra are isotropic for light polarization along the in-plane directions. Figure \ref{4}(a) shows the imaginary and real parts of the dielectric function of MoSi$_{2}$N$_{4}$ monolayer.
The first absorption peak of Im ($\varepsilon$) appears at 2.28 eV which is in the visible range of light. The static dielectric constant (real part of the dielectric constant at zero energy) for MoSi$_{2}$N$_{4}$ single-layer is 4.90. 
In Drude model the plasma frequencies are defined by the roots of Re ($\varepsilon$) with x = 0 line \cite{masoud1,masoud2}. 
The first plasma frequency of MoSi$_{2}$N$_{4}$ monolayer appears at 4.99 eV which is related to the $\pi$ electron plasmon peak. Figure \ref{4}(b) demonstrates the absorption coefficient $\alpha$ of MoSi$_{2}$N$_{4}$ single-layer in-plane polarization. The first absorption peak appears 540 nm (2.31 eV) which is in visible range which is a good agreement with the experimental results \cite{Hong2020}. MoSi$_{2}$N$_{4}$ monolayer has the ability to absorb a wide range of the visible light radiation due to its suitable bandgap. MoSi$_{2}$N$_{4}$ can be lucrative material in optoelectronic applications.

\section{Thermal properties}

\begin{figure*}[!htb]
	\includegraphics[scale=1.5]{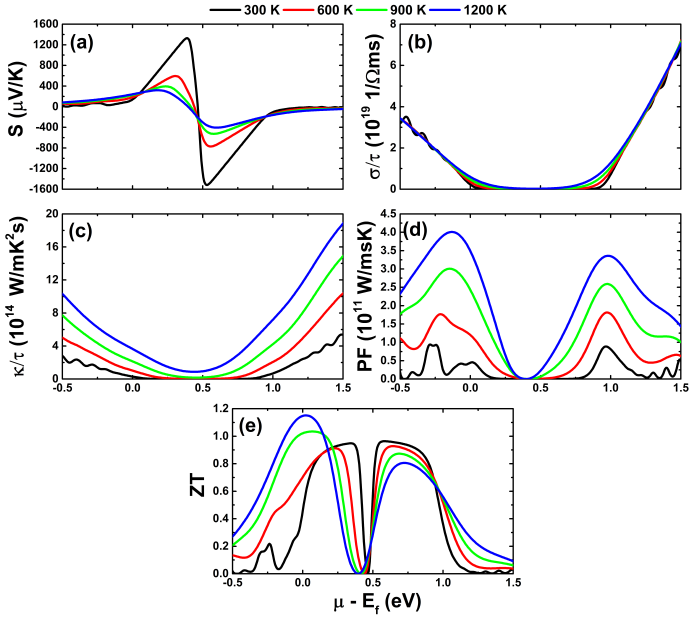}
	\caption{(a) Seebeck coefficient, (b) Electrical conductivity, (c) Thermal electronic conductivity, (d) Power factor and (e) Figure of merit of MoSi$_{2}$N$_{4}$ monolayer as a function of chemical potential.}
	\label{5}
\end{figure*}

The thermoelectric potential is an important property that can be studied through the Seebeck coefficient (thermopower). 
Materials with large Seebeck coefficient possess good capability of pushing electrons from the hot to the cold regions. 
Thermoelectric properties of the MoSi$_{2}$N$_{4}$ monolayer is calculated using the semiclassical Boltzmann transport theory in the rigid band approximation and constant scattering time approximation framework \cite{madsen2006boltztrap}. 
As a first step, the transport distribution tensors $\sigma_{\alpha \beta}(\epsilon)$ is calculated via interpolation of the electronic band structure by the following expression:

\begin{equation}
	\sigma_{\alpha \beta}(\epsilon) = \frac{e^{2}}{N}\sum_{i,k} \tau_{i,k}\nu_{\alpha}(i,k)\nu_{\beta}(i,k)\frac{\partial(\epsilon - \epsilon_{i,k})}{\partial \epsilon},
	\end{equation}
	herein, $\nu(i,k)$ is the group velocity component with tensor indices $\alpha$ and $\beta$, $N$ refers to the k-point number and $\tau$ denotes relaxation time. Then the Seebeck coefficient, electrical conductivity and electronic thermal conductivity are deduced from $\sigma_{\alpha \beta}(\epsilon)$ as follow: 
	\begin{equation}
	\sigma_{\alpha \beta}(T,\mu) = \frac{1}{\Omega}\int \sigma_{\alpha \beta}(\epsilon)\left[-\frac{\partial f_{0}(T,\epsilon, \mu)}{\partial \epsilon}\right] d\epsilon,
	\end{equation}
	\begin{equation}
	S_{\alpha \beta}(T,\mu) = \frac{1}{eT\Omega \sigma_{\alpha \beta}(T,\mu)}\int \sigma_{\alpha \beta}(\epsilon)(\epsilon - \mu)\left[-\frac{\partial f_{0}(T,\epsilon, \mu)}{\partial \epsilon}\right] d\epsilon,
	\end{equation}
	\begin{equation}
	\kappa^{0}_{\alpha \beta}(T,\mu) = \frac{1}{e^{2}T\Omega}\int \sigma_{\alpha \beta}(\epsilon)(\epsilon - \mu)^{2}\left[-\frac{\partial f_{0}(T,\epsilon, \mu)}{\partial \epsilon}\right] d\epsilon.
	\end{equation}

Figure \ref{5}(a) displays the MoSi$_{2}$N$_{4}$ monolayer Seebeck coefficient ($S$) as a function of chemical potential at different temperatures. One can see that the increasing temperature decreases the $S$ in the chemical potential range 0.1 eV to 1 eV with a small shift of peaks towards the low chemical potential. The $S$ drops rapidly to negligible values outside the previous the chemical potential range, suggesting that good thermoelectric properties of MoSi$_{2}$N$_{4}$ monolayer in this chemical potentials range.\\
Electrical conductivity measures the flow of the charge carriers in free movement from the hot to cold regions. Thermoelectric materials are expected to have small electrical resistance, that is large electrical conductivity to avoid the energy loss generated by the heating effects. Figure \ref{5}(b) shows the electrical conductivity of MoSi$_{2}$N$_{4}$ monolayer. The electrical conductivity is quite small in the chemical potential range from 0.25 eV to 0.62 eV, and it slightly increases with increasing the temperature. Beyond -0.13 eV and 1.13 eV, $\sigma$  nearly linearly increases with the chemical potential, in which the temperature has no significant effect. Within the thermoelectric materials, the temperature gradient should be maintained, therefore small thermal conductivity is a desirable feature. 

Figure \ref{5}(c) shows the thermal conductivity as a function of chemical potential using the Wiedemann-Franz law, $\kappa = \sigma LT$, where $L$ and $T$ refer to the Lorenz number and temperature. The thermal conductivity  behavior is similar to the electrical conductivity behavior (Fig. \ref{4}(b)).
  
The temperature change induces a considerable variation of the electronic thermal conductivity. Specifically, at a given chemical potential, $k_{el}$ increases as the temperature increases. For example, at the considered lower chemical potential limit, $k_{el}$ is  2.48 and 10.97 (10$^{14}$ W/msK$^{2}$), which corresponding to an increase of the order of 342.34\%, when raising the temperature from 300 to 1200 K. At the upper limit, the increasing order is of 258.06\% (from 5.46 to 19.55 (10$^{14}$ W/msK$^{2}$)). 

Good capability of producing electricity can be obtained with the high thermoelectric potential and high electrical conductivity. To measure this property, the power factor has been proposed, which can be determined as follows: $PF = S^{2} \sigma$. Figure \ref{5}(d) illustrates the power factor of MoSi$_{2}$N$_{4}$ single layer as a function of chemical potential at different temperatures. The heat conversion into electricity may be favored by the temperature since the $PF$ increases considerably according to increase temperature. The maximum values of $PF$ are located at 0.12 eV and 0.98 eV. It appears that the power factor is larger in the negative chemical potential region than the positive region, and the difference becomes clearer at high temperatures.\\ 
The thermoelectric performance of materials can be studied using the dimensionless figure of merit, $ZT = \frac{S^{2} \sigma T}{\kappa}$. As the figure of merit increases the thermoelectric performance increases. Figure \ref{5}(e) shows the dependence of MoSi$_{2}$N$_{4}$ monolayer $ZT$ on the chemical potential. In the chemical potential range from 0.35 eV to 0.5 eV, $ZT$ has small values (lower than
0.1). Beyond this range $ZT$ increases rapidly to reach the largest
values (0.8-1.12) and it depends on the temperature. In the chemical potential range from -0.45 eV to 0.5 eV, $ZT$ shifts towards the low potential and, in the same time, the maximum value increases as the temperature increases. However, above the previous chemical potential range, $ZT$ shifts towards the high potetial  and the maximum value decreases as the temperature increases. It is worth mentioning that the figure of merit is close to unity, suggesting that the MoSi$_{2}$N$_{4}$ single layer may be a prospective candidate for thermoelectric nano-devices. 

\section{Photocatalytic properties}

\begin{figure}
	\includegraphics[scale=1]{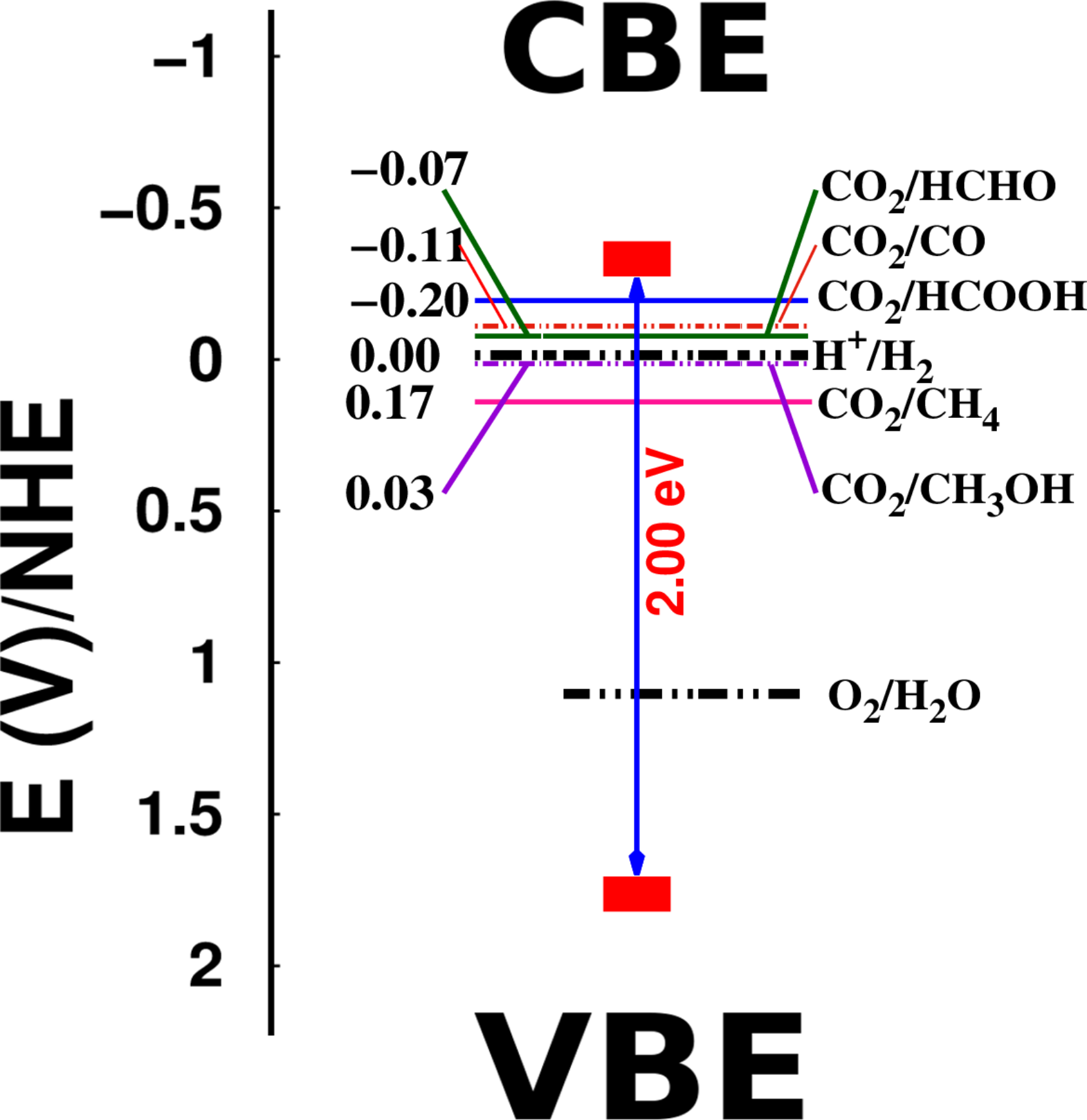}
	\caption{Band alignments of MoSi$_{2}$N$_{4}$ monolayer for photocatalytic water splitting and carbon dioxide reduction The band edges are given with respect to the NHE (normal hydrogen electrode) potential (in Volts).}
	\label{6}
\end{figure}

Most importantly, MoSi$_{2}$N$_{4}$ exhibits a suitable band gap $\sim$ 2 eV and the band edges CBE and VBE must be higher (more negative) and lower (more positive) than the hydrogen reduction potential of H$_{+}$/H$_{2}$ and the water oxidation potential of H$_{2}$O/O$_{2}$, respectively, for water splitting. For the band edges for CO$_{2}$ reduction, CBM must be higher than the CO$_{2}$ reduction of natural gaes and the VBM has the same conduction in the water splitting process. The CBE is computed from the relation $E_{CBE} = X - 0.5 E_{gap} - 4.5$ eV \cite{A1,A2,A3}, then the valence band edge is calculated from $E_{VBE} = E_{CBE} + E_{gap}$, where $X$ is the geometric mean of the electro-negativity of the ingredient atoms \cite{A4} and $4.5$ eV is is the free energy of the electron (with respect to the vacuum level). Figure \ref{6} illustrates the alignment of oxidation and reduction potentials for water splitting and CO$_{2}$ reductions with respect to the band edges of MoSi$_{2}$N$_{4}$ which suggest that it is potential candidates for photocatalytic water splitting and CO$_{2}$ reduction.

\section{Conclusion}

In this study, the stability and the physical properties of MoSi$_{2}$N$_{4}$ single layer were investigated by hybrid density functional theory. The stability was studied by the dynamical method in terms of phonon dispersion relation and by mechanical method in terms of elastic constant parameters, Young's modulus and  Poisson's ratio. There were no negative values in the phonon dispersion, the elastic parameters satisfied the criteria of stability, and Young's modulus and  Poisson's ratio were compared to the corresponding graphene values. The thermoelectric properties indicated that large thermopower can be obtained in the chemical range from 0 to 1 eV, and this parameter exhibited the opposite chemical potential-dependence as compared with the electrical conductivity. The figure of merit values suggested promising thermoelectric applicability of the MoSi$_{2}$N$_{4}$ monolayer, and the performance can be enhanced by increasing temperature. The linear photon energy-dependent optical response of MoSi$_{2}$N$_{4}$ monolayer, investigated in terms of RPA+HSE06 approach. Our results indicated that the absorption peaks of along in-plane polarization are located in the visible range of light, suggesting its prospect for applications in optoelectronics and nanoelectronics. Moreover, MoSi$_{2}$N$_{4}$ layer can be a good candidate photocatalytic for water splitting and CO$_{2}$ reduction due to its band edges positions and suitable bandgap. 
 
\section{Conflicts of interest}
The authors declare that there are no conflicts of interest regarding the publication of this paper.

\section{ACKNOWLEDGMENTS}
This work was supported by the National Research Foundation of Korea(NRF) grant funded by the Korea government(MSIT)(NRF-2017R1A2B2011989). 
Computational resources were provided by the Flemish Supercomputer Center (VSC) and TUBITAK ULAKBIM, High Performance and Grid Computing Center (Tr-Grid e-Infrastructure).

\bibliography{ref}

\end{document}